\documentstyle[12pt,epsf]{article}
\begin{document}
\setlength{\baselineskip}{24pt}
\setlength{\textheight}{24cm}
\setlength{\textwidth}{20cm}

\title{Power law fluctuation generator based on analog electrical circuit }
\author{Aki-Hiro Sato$^{1}$\footnote{Electronic
mail:aki@sawada.riec.tohoku.ac.jp}, Hideki Takayasu$^{2}$ and Yasuji
Sawada$^{1}$ \\ 
$^{1}$Research Institute of Electrical Communication, \\
Tohoku University, Sendai 980-8577, Japan.\\
$^{2}$ Sony Computer Science Lab., Takanawa Muse Bldg., 3-14-13, \\
Higashi-Gotanda, Shinagawa-ku, Tokyo 141-0022, Japan.
}
\date{\today}

\maketitle

\begin{center}
{\bf Abstract}
\end{center}
{
We constructed an analog circuit generating fluctuations of which a
probability density function has power law tails. In the
circuit fluctuations with an arbitrary exponent of the power law can be 
obtained by tuning the resistance. A theory of a differential equation 
with both multiplicative and additive noises which describes the circuit 
is introduced. The circuit is composed of a noise generator, an analog 
multiplier and an integral circuit. Sequential outputs of the
circuit are observed and their probability density function and
autocorrelation coefficients are shown. It is found that correlation
time of the autocorrelation coefficient is dependent on the power law
exponent.  
}

\section{Introduction}
Random multiplicative process (RMP) is attracting much interest
recently as a new mechanism of generating power law fluctuations.
The process is described by a Langevin system with both
multiplicative and additive noises. It was proved that a PDF for 
its amplitude has power law tails when the noises
are independent\cite{Takayasu,Jogi,Sornette}. The continuous time 
version of RMP is described by a stochastic differential equation with 
both multiplicative and additive noises. It was shown that a stationary
probability density function for its amplitude has power law tails 
when the noises are Gaussian white noise\cite{Deutsch,Venkataramani,Nakao}.
This type of equations has been studied in chaos on-off
intermittency\cite{Intermittency}, motions of a polymer in turbulent 
fluid\cite{Polymer}, nonlinear coupled oscillators\cite{Kuramoto} and
price changes in economics\cite{Aki-Stock}.

In this paper we propose an electrical generator of random
fluctuations based on the theory of RMP for the power law 
fluctuations by introducing an analog electrical circuit 
consisted of operational amplifiers and an analog multiplier. 

The paper is organized as follows. In Sec. \ref{sec:Theory} we
make a brief explanation of the theory of the continuous time RMP.
In Sec. \ref{sec:Circuit} we show the circuit diagram of the proposed
generator having power law fluctuations and explain its mechanism.
In Sec. \ref{sec:ResultAndDiscussion} we show the data of sequential
outputs, their probability distribution function (PDF) and their 
autocorrelation coefficients with discussion. Sec. \ref{sec:Conclusion} 
is devoted for concluding remarks. 

\section{Theory of RMP}
\label{sec:Theory}
We here review a brief outline of the stochastic differential equation
with both multiplicative and additive
noises\cite{Deutsch,Venkataramani,Nakao}. The stochastic differential 
equation is given by
\begin{equation}
\frac{dv}{dt} = \nu(t)v(t) + \xi(t),
\label{eq:RMP}
\end{equation}
where $v(t)$ is a dynamical variable, and $\nu(t)$ and $\xi(t)$
represent a multiplicative noise and an additive noise, respectively.
Representing ensemble average by $\langle \ldots \rangle$ we require
$\langle \nu(t) \rangle = \bar{\nu}$, and $\langle \xi(t) \rangle = 0$. 
When $\nu(t)$ and $\xi(t)$ follow Gaussian white noises with their
strengths characterized by $D_{\nu}$ and $D_{\xi}$, respectively,
\begin{eqnarray}
\langle [\nu(t_1) - \bar{\nu}][\nu(t_2) - \bar{\nu}]\rangle 
&=& 2D_{\nu}\delta(t_1 - t_2),
\\
\langle \xi(t_1) \xi(t_2) \rangle &=& 2D_{\xi}\delta(t_1 - t_2),
\end{eqnarray}
the PDF for the dynamical variable $v$ in eq. (\ref{eq:RMP}),
$p(v,t)$, is known to follow the generalized Fokker-Planck equation,
\begin{equation}
\frac{\partial}{\partial t}p(v,t) = \frac{\partial}{\partial
v}\Bigl[-(\bar{\nu}+ D_{\nu})v + \frac{\partial}{\partial
v}(D_{\nu}v^2 + D_{\xi})\Bigr]p(v,t). 
\label{eq:Fokker-Planck}
\end{equation}
A steady state solution corresponding to a zero-current state can be
obtained by putting the term of $\frac{\partial}{\partial v}$ to 0. 
By solving eq. (\ref{eq:Fokker-Planck}) we get a stationary distribution,
\begin{equation}
p(v) \propto (D_{\xi}+D_{\nu}v^2)^{\frac{\bar{\nu}}{2D_{\nu}}-\frac{1}{2}},
\label{eq:steady-state-pdf}
\end{equation}
which has power law tails for large $v$, 
\begin{equation}
p(v) \propto |v|^{\bar{\nu}/D_{\nu}-1}.
\label{eq:power-law}
\end{equation}
For convenience of notation we define an exponent of the power law
tail as 
\begin{equation}
p(v) \propto |v|^{-\beta-1}.
\label{eq:define-of-power-law-exponent}
\end{equation}
Comparing eq. (\ref{eq:power-law}) with
eq. (\ref{eq:define-of-power-law-exponent}) yields 
\begin{equation}
\beta = -\frac{\bar{\nu}}{D_{\nu}}.
\label{eq:beta}
\end{equation}
It is useful to introduce a cumulative distribution function (CDF)
defined as 
\begin{equation}
P(\geq|v|)=\int_{-\infty}^{-|v|}p(v')dv' + \int_{|v|}^{\infty}p(v')dv'.
\label{eq:CDF}
\end{equation}
Then the CDF of $v$ in eq. (\ref{eq:RMP}) is given as 
\begin{equation}
P(\geq|v|) \propto |v|^{-\beta},
\end{equation}
namely, the slope in log-log plots directly gives the power law
exponent $\beta$. The corresponding PDF of
eq. (\ref{eq:steady-state-pdf}) is given as
\begin{equation}
p(v) \propto
\Bigl(1+\frac{v^2}{s^2}\Bigr)^{-\beta/2-1/2},
\label{eq:Levy}
\end{equation}
where $s=\sqrt{\frac{D_{\xi}}{D_{\nu}}}$. Namely $p(v)$ has power law 
tails for $v>s$.

\section{The circuit}
\label{sec:Circuit}
From the assumption for the multiplicative noise in the above theory
it is important that $\nu(t)$ takes both positive and negative values
to realize the power law tails. In an electrical circuit this means that
it is necessary for the circuit to include both positive and negative
feedbacks. We solve this problem by using an analog multiplier. A
block diagram of our circuit and an implemented noise generator are 
shown in fig.\ref{fig:circuit}. In the figures $v_o$ represents the
output voltage, and $\mu(t)$ is the output directly from the noise
generator. As seen from this figure the circuit contains the noise
generator, an analog multiplier (Analog devices, 10MHz, 4-quadrant)
and an operational amplifier (National Semiconductor, LF157) for
integrator. 

The output of the noise generator plays a role of the multiplicative
noise in eq. (\ref{eq:RMP}). In the noise generator a shot noise
between the zener diode in fig. \ref{fig:circuit} is amplified by an
operational amplifier, it passes through a high pass filter to be taken
out from the circuit. LF157 has the 20M product of a voltage gain (G) 
and a bandwidth (B). The bandwidth is given by B=200kHz in the noise 
generator because we put G=100. Thus we expect a frequency
characteristics of the noise generator to be up to 200kHz . We
perform the positive and negative feedback by multiplying $v_o$ with the
output of the noise generator $\mu(t)$ in the analog multiplier and by
connecting the product to the negative input with the operational
amplifier for integrator.   
Although the additive noise term is not explicitly added in the
circuit, it derives from either thermal noises of the operational
amplifier or from an external electro-magnetic noises.

An equation equivalent to the circuit diagram of
fig. \ref{fig:circuit} is given as
\begin{equation}
\frac{dv_o}{dt} 
= \Bigl(\frac{1}{R_{f}C}+\frac{k}{R_{v}C}\mu(t)\Bigr)v_o + \xi(t),
\label{eq:circuit}
\end{equation}
where $k$ is a factor of the multiplier with $k=1/10$, and
$\xi(t)$ represents the additive noise effect. The strength of multiplicative
noise $\mu(t)$ depends on the value of a variable resistor $R_v$ in
fig. \ref{fig:circuit} because $R_v$ is a factor of $\mu(t)$ in
eq. (\ref{eq:circuit}). Therefore, we expect that the power law exponent for
the output is  a function of $R_v$ like eq. (\ref{eq:beta}). For the
output of the noise generator $\mu(t)$ we show a PDF and an
autocorrelation coefficient in figs. \ref{fig:multiplicative-pdf} and
\ref{fig:multiplicative-acorr}, respectively. The autocorrelation
coefficient is defined as 
\begin{equation}
K_1(\tau) = 
\frac{\langle \mu(t+\tau) \mu(t) \rangle - \langle \mu(t+\tau) \rangle
\langle \mu(t) \rangle}{\langle \mu(t)^2 \rangle - \langle \mu(t) \rangle^2},
\label{eq:K1}
\end{equation}
We sampled outputs of the noise generator through 12bit AD-converter
(Microscience, ADM-652AT) and processed them in a computer. A sampling
frequency is 125kHz. We find that the PDF can be approximated by the
Gaussian distribution and that the autocorrelation coefficient 
decays quickly. This result means that the frequency characteristics of
the noise generator is up to around 200kHz. Therefore we consider $\mu(t)$
as a Gaussian white noise under 200kHz.

\section{Results and Discussions}
\label{sec:ResultAndDiscussion}
We observe the output of the circuit $v_o$ in respect of $R_v$ since 
we expect that the power law exponent $\beta$ is a function of $R_v$ 
from the theory. We detected the output through a 12-bit AD
converter and processed it as digital data. A sampling frequency is
125kHz throughout the observations. Here, we explain the method of
adjusting an offset of the operational amplifier for integrator. 
We adjust the variable resistor added to the operational amplifier
for a temporal average of output $\langle v \rangle$ to be almost 
likely zero when the value of $R_v$  is the smallest. As shown in 
fig. \ref{fig:error} the reason is that the temporal  
average shows exponential decay for $R_v$. We keep the zero-average of
$v(t)$ throughout all observations in this way.

We show examples of temporal output at $R_v=5$ and $R_v=150$ in fig. 
\ref{fig:timeseries}. We find that amplitude of the output $v(t)$
depends on the value of $R_v$. The PDF for the temporal output on 
$R_v=5\Omega$ and $R_v=150\Omega$ are shown in fig.\ref{fig:pdf}. 
The tail's form of the PDF depends on the value of $R_v$.
To see the difference between the positive and negative tails in
detail we show log-log plots of the PDF in the positive and negative
domains in fig. \ref{fig:pdf-pn}. 
It is obvious that the slope of the tail in the positive domain looks 
different from that in the negative domain. There are tow possible 
reasons on this effect. One is that an average of the additive noise is not
completely zero. The other is that a distribution of the additive
noise is asymmetrical.  

To roughly estimate the dependence of the exponent $\beta$ on $R_v$ we 
show log-log plots of CDF of $v(t)$ at $R_v=5$,$25$,$50$,$100$,$150\Omega$ 
in fig. \ref{fig:cdfs}. Log-log plots of CDF have a liner part for
about a decade. We clearly find that the exponent $\beta$ depends on
$R_v$ as expected qualitatively. However, we can not apply
eq. (\ref{eq:beta}) directly to these experimental results 
because the multiplicative noise generated by the noise generator is not 
an ideal white noise but is correlated at some frequencies. The authors 
have proved a relation to the power law exponent of RMP with
correlated multiplicative noise in a discrete time
version\cite{Aki-Colored}, yet, in the continuous time version the
theoretical relation between the exponent and the correlated
multiplicative noise is an interesting open problem. 

From the bending points of CDFs in fig.\ref{fig:cdfs} we can estimate
$s \approx 0.1$ for eq. (\ref{eq:Levy}). Namely, the strength of the
additive noise is $10^{-2}$ times as large as that of the
multiplicative noise.  

The autocorrelation coefficient of outputs, defined in eq. (\ref{eq:K1}),
are shown in fig. \ref{fig:acorr}. A correlated time of the autocorrelation 
coefficient $K_1(\tau)$ on $R_v=5\Omega$ is longer than the case of
$R_v=150\Omega$. This implies that the correlation time depends on the
value of $R_v$. In economics it is often more interesting to observe the
autocorrelation coefficient of squared outputs than the standard
autocorrelation coefficient\cite{Volatility}. The autocorrelation
coefficient of squared outputs called the volatility correlation is
defined as 
\begin{equation}
K_2(\tau) = \frac{\langle v(t+\tau)^2 v(t)^2 \rangle - \langle v(t+\tau)^2
\rangle \langle v(t)^2 \rangle}{\langle v(t)^4 \rangle - \langle v(t)^2 
\rangle^2}.
\end{equation}
As shown in fig. \ref{fig:acorr2} the correlation time of the
autocorrelation coefficient $K_2(\tau)$ also depend on the power law
exponent. The reason why $K_1(\tau)$ or $K_2(\tau)$ have longer
correlation time for smaller value of $R_v$ can be simply answered in
the following way. Comparing the time series in
fig. \ref{fig:timeseries} a and b we find that the motions are
basically characterized by repeating the processes of 
starting from nearby zero, reaching the peak voltage and
returning to zero again. As the maximum peak levels for $R_v=5\Omega$
are roughly several times larger than the case of $R_v=150\Omega$,
each process of going up and down roughly takes several times longer
time leading the longer correlation time.

\section{Conclusion}
\label{sec:Conclusion}
We proposed an electrical analog circuit generating the fluctuation
of which a probability density function has the power law tails. The
circuit includes a noise generator, an analog multiplier and an 
operational amplifier for integrator. In our proposal circuit fluctuations
with an arbitrary exponent of the power law can be obtained by tuning
the value of resistance. We observed temporal outputs of the circuit
and calculated the probability density function, the autocorrelation
coefficient and the autocorrelation coefficient of the squared
outputs. We show the probability density function 
with power law tails. The probability density function has
different tails in positive and negative domains because 
the average of the additive noise is non-zero or its distribution 
function is asymmetrical. We conclude that the correlation time of the
autocorrelation coefficient and that of the squared outputs depend on
the value of resistance in the circuit. We expect that  
the proposed circuit is applicable to generate fluctuations having
power law distribution in a much cheaper way than any digital
computing methods. Moreover fluctuations of our circuit may be 
of use for risk estimation in foreign exchange or stock market in the
near future.

\section{Acknowledgment} 
One of the authors (A.-H. Sato) wishes to thank Yoshihiro Hayakawa 
for his stimulative discussion and his lecture on electrical circuits.

%======================================================================
\begin{figure}[h]
%\epsfile{file=Circuit.ps,scale=0.5}
\caption{Block diagram of all circuits with a noise generator. The
circuit contains a noise generator, a analog multiplier and
an operational amplifier for integrator. $R_f=100$k$\Omega$,
$C=10$pF,$R_v=200\Omega$. A variable resistor under the operational 
amplifier is for adjustment of the offset.}
\label{fig:circuit}
\end{figure}
%======================================================================
%======================================================================
\begin{figure}[h]
%\epsfile{file=multiplicative_pdf.ps,scale=0.5}
\caption{Semi-log plots of the probability density function of the 
multiplicative noise $\mu(t)$. Filled circles represent observation. A
solid curve represents a Gaussian distribution with the same average
and deviation as observation. From numerical estimation the average
is 0.005103, and the variance 0.006846.} 
\label{fig:multiplicative-pdf}
\end{figure}
%======================================================================
%======================================================================
\begin{figure}[h]
%\epsfile{file=multiplicative_acorr.ps,scale=0.5}
\caption{Autocorrelation coefficient of the multiplicative noise $\mu(t)$
for 125kHz sampling frequency. Filled circles represent observation.}
\label{fig:multiplicative-acorr}
\end{figure}
%======================================================================
%======================================================================
\begin{figure}[h]
%\epsfile{file=offset.ps,scale=0.5}
\caption{Semi-log plots of relation between a temporal average of the 
output $\langle v \rangle$ and the value of $R_v$. Filled and Unfilled 
circles represent observation with different offset voltage in the operational 
amplifier. A solid line represents $\langle v \rangle = 4.5\exp(-0.012 R_v)$.
The fit of the data to exponential is good from 0 to 150$\Omega$.}
\label{fig:error}
\end{figure}
%======================================================================
%======================================================================
\begin{figure}[h]
%\epsfile{file=seq5.ps,scale=0.5}
%\epsfile{file=seq150.ps,scale=0.5}
\caption{Signal outputs for $R_v=5\Omega$ (a) and $R_v=150\Omega$ (b).
The amplitude for $R_v=5\Omega$ is larger than for $R_v=150\Omega$.}
\label{fig:timeseries}
\end{figure}
%======================================================================
%======================================================================
\begin{figure}[h]
%\epsfile{file=pdfs.ps,scale=0.5}
\caption{Semi-log plots of probability density functions $R_v=5\Omega$ 
(unfilled circles) and $R_v=150\Omega$ (filled circles). The probability 
density function for $R_v=5\Omega$ has wider tails than for $R_v=150\Omega$.}
\label{fig:pdf}
\end{figure}
%======================================================================
%======================================================================
\begin{figure}[h]
%\epsfile{file=pdf-pn5.ps,scale=0.5}
\caption{Log-log plots of the probability density function for $R_v=5\Omega$.
Unfilled circles represent it in the positive domain and filled circles in 
the negative.}
\label{fig:pdf-pn}
\end{figure}
%======================================================================
%======================================================================
\begin{figure}[h]
%\epsfile{file=cdfs.ps,scale=0.5}
\caption{Log-log plots of CDF of the output $v_o(t)$ at 
$R_v=5,25,50,100,150\Omega$. CDFs at $R_v=50,100$ show 
straight lines with different slopes between a decade.}
\label{fig:cdfs}
\end{figure}
%======================================================================
%======================================================================
\begin{figure}[h]
%\epsfile{file=acorr.ps,scale=0.5}
\caption{Log-log plots of autocorrelation coefficients for $R_v=5\Omega$
(unfilled circle) and $R_v=150\Omega$ (filled circle). The autocorrelation
coefficient for $R_v=5\Omega$ has a longer decay time than for 
$R_v=150\Omega$.}
\label{fig:acorr}
\end{figure}
%======================================================================
%======================================================================
\begin{figure}[h]
%\epsfile{file=acorr2.ps,scale=0.5}
\caption{Log-log plots of autocorrelation coefficients for squared output.
Unfilled circles represents for $R_v=5\Omega$, and filled circles for 
$R_v=150\Omega$. }
\label{fig:acorr2}
\end{figure}
%======================================================================

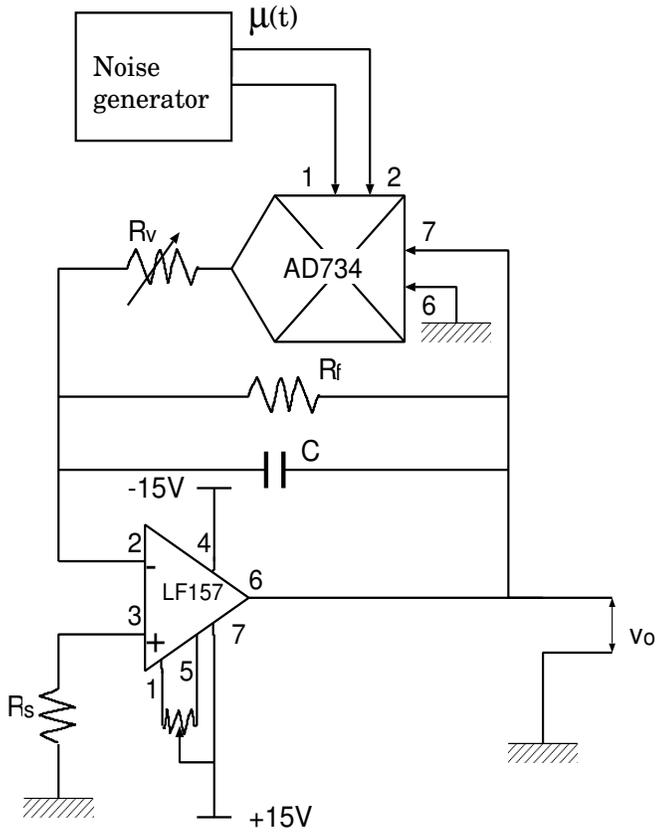
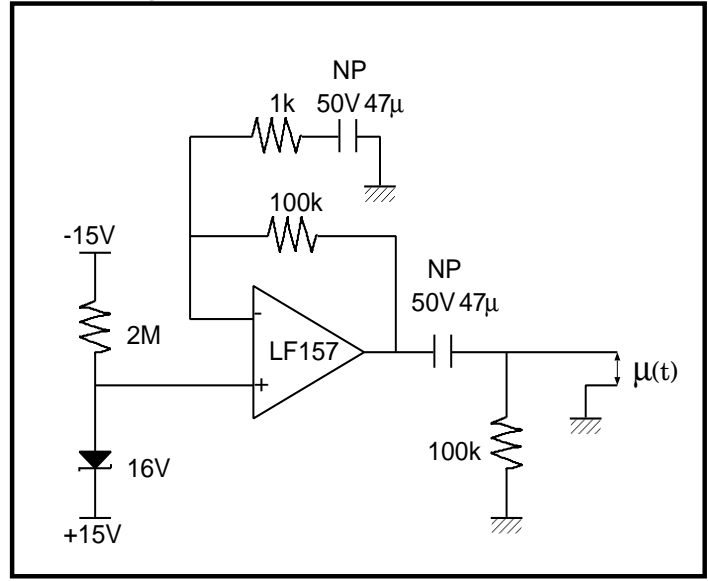
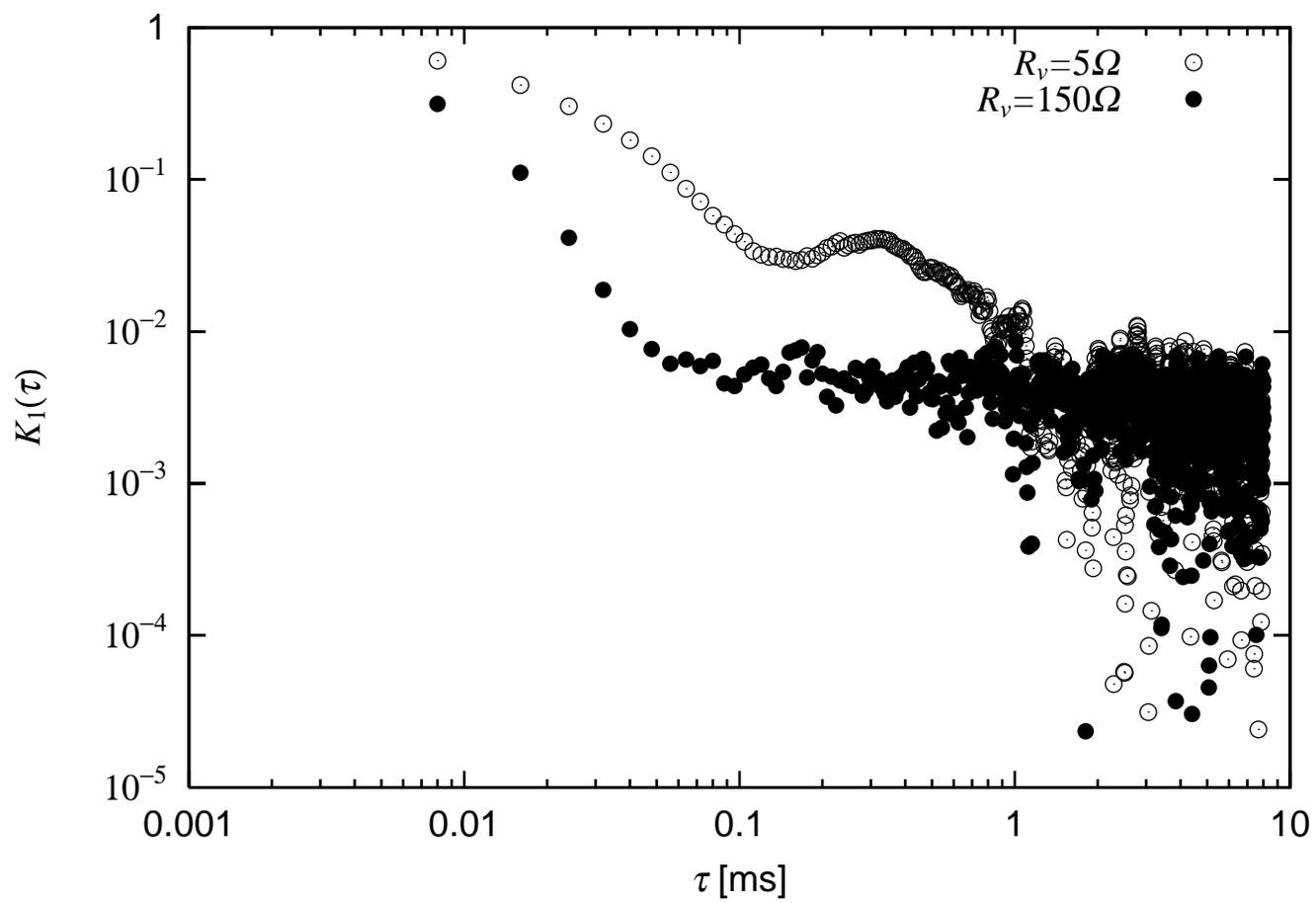
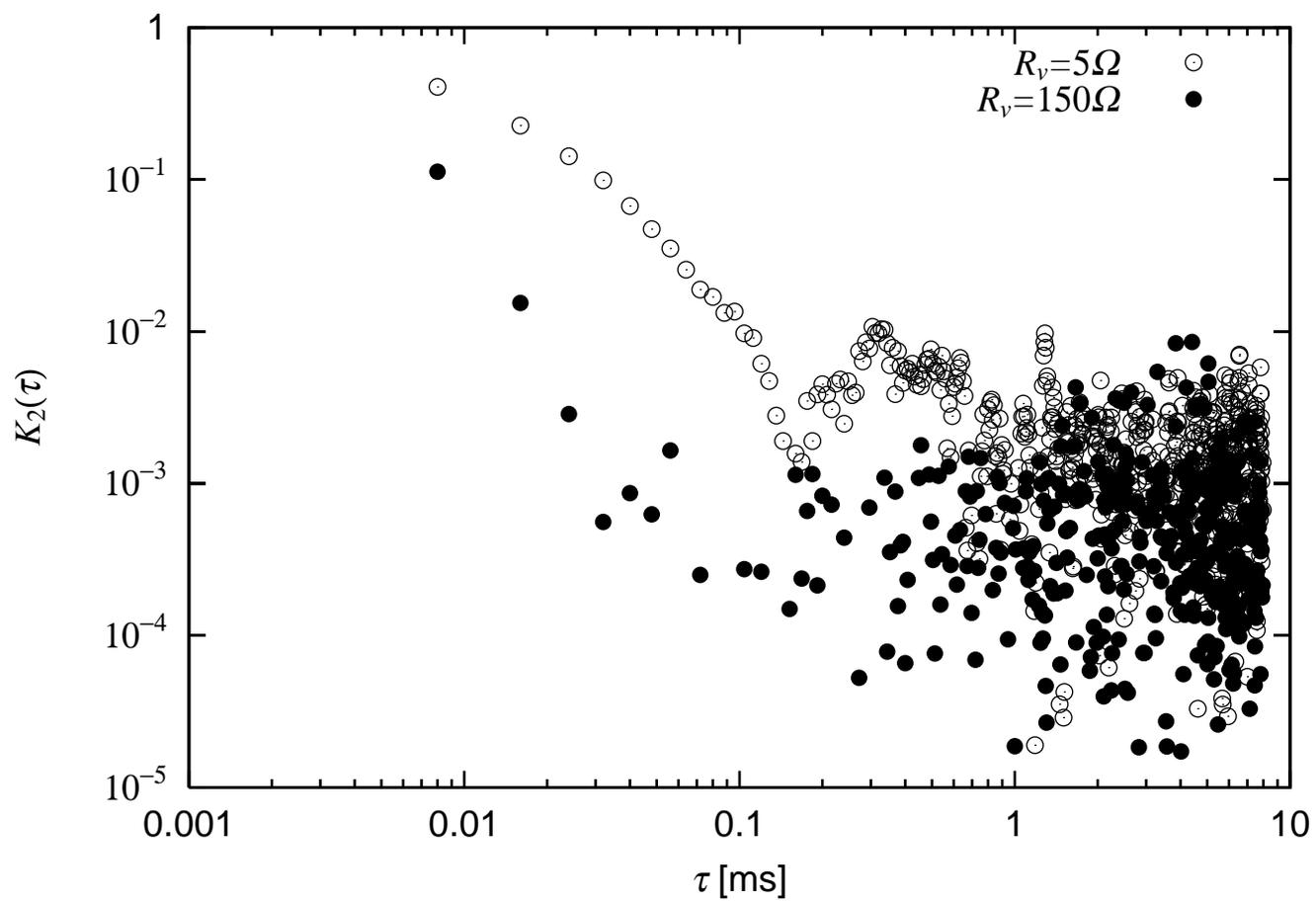
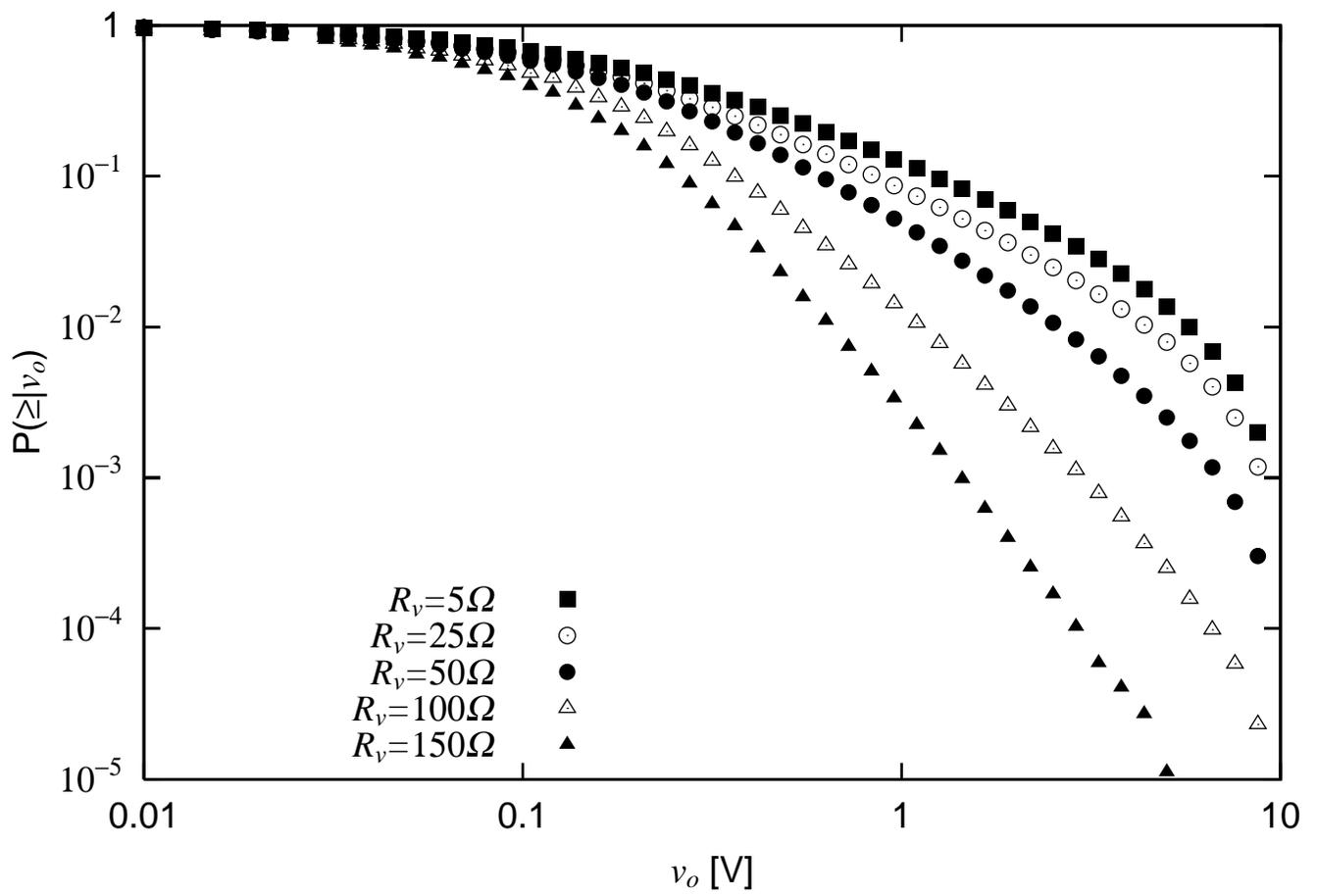
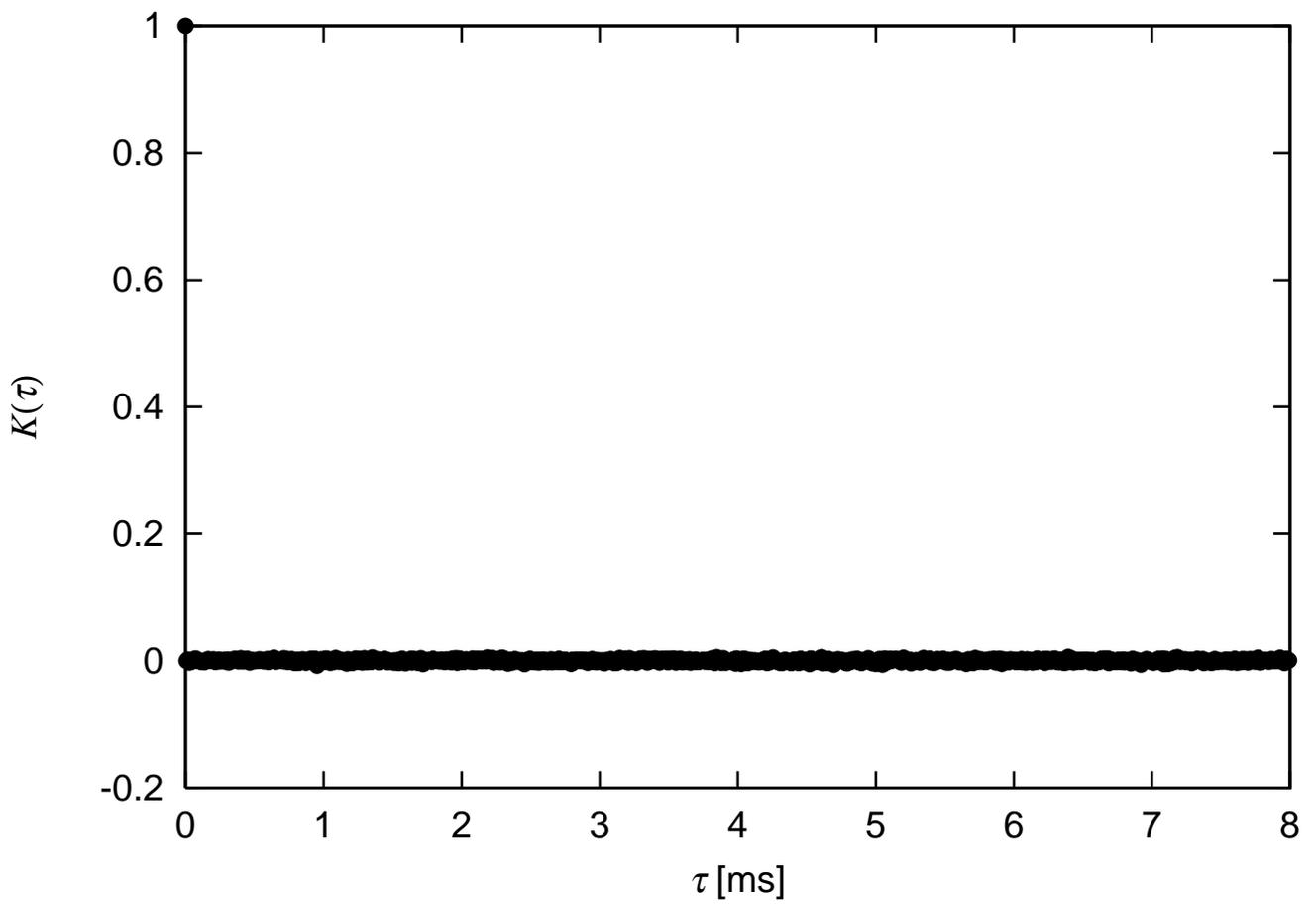
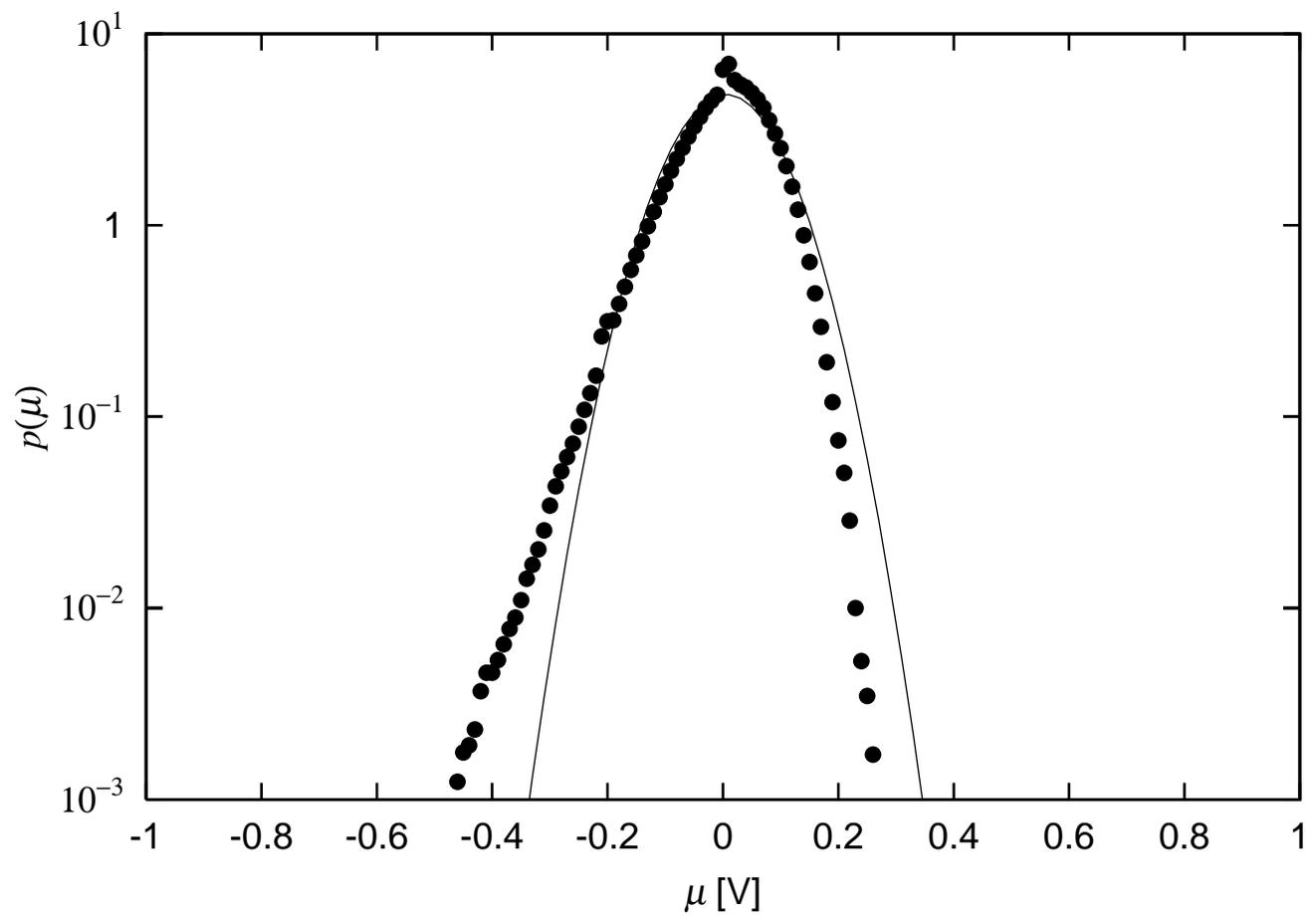
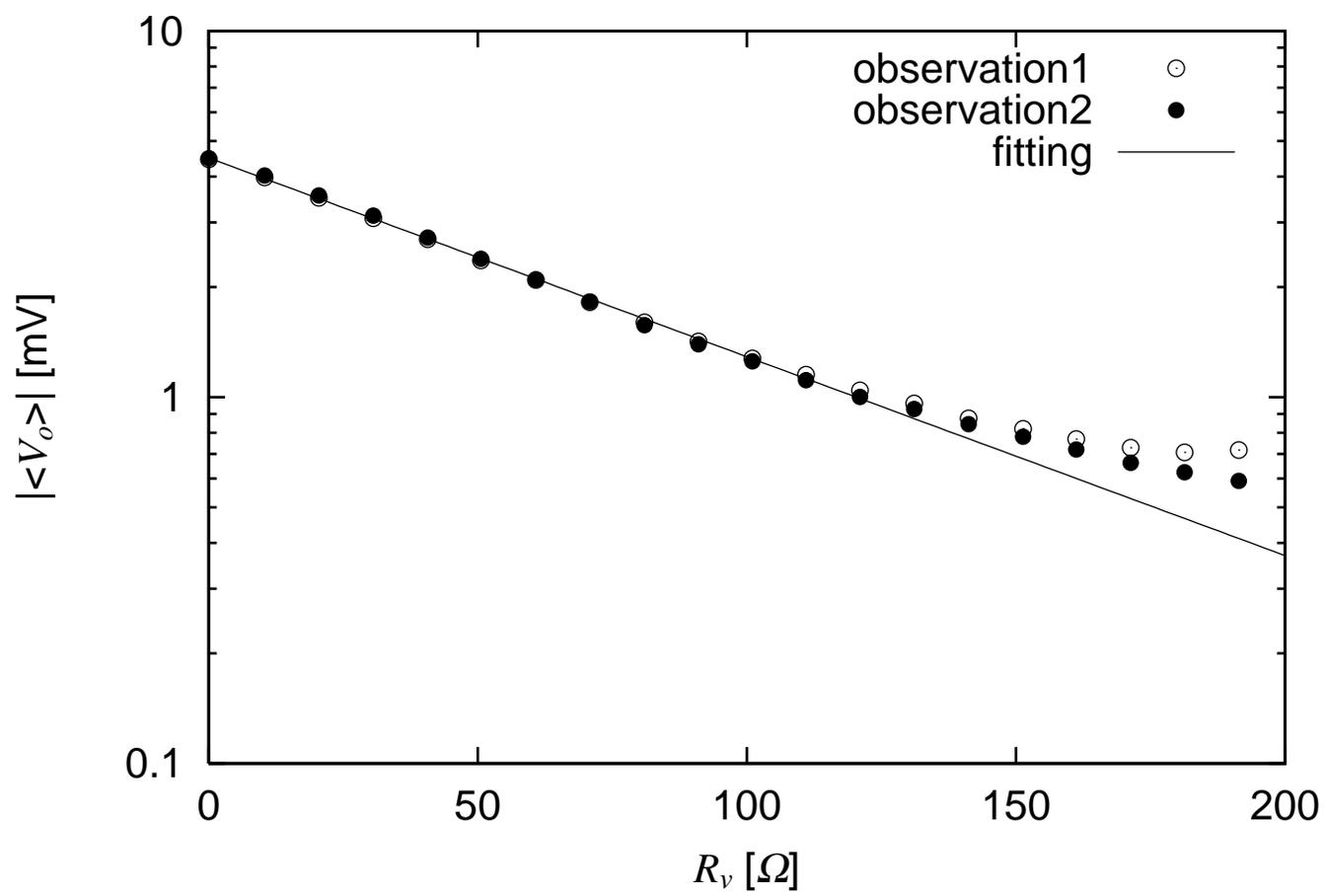
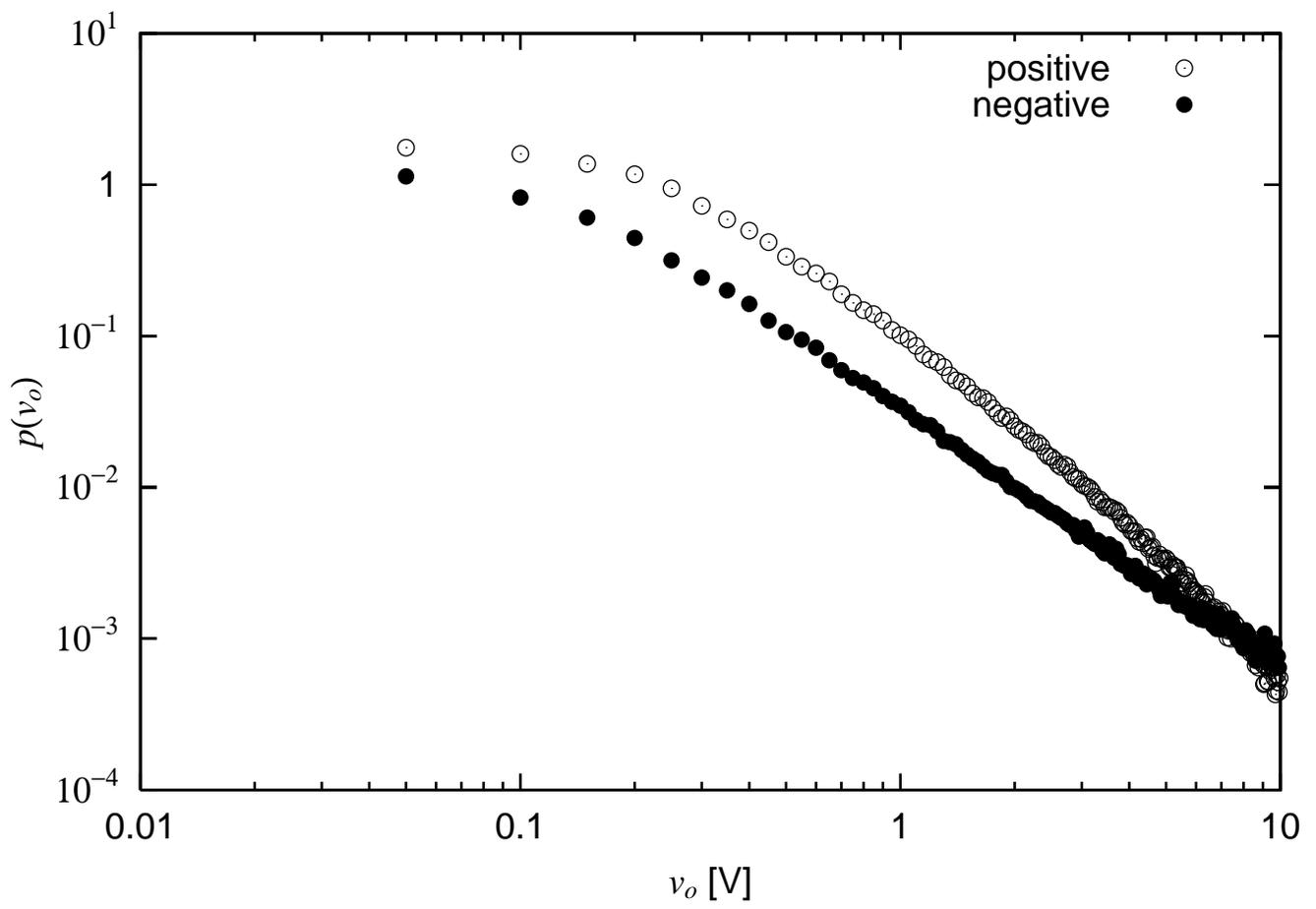
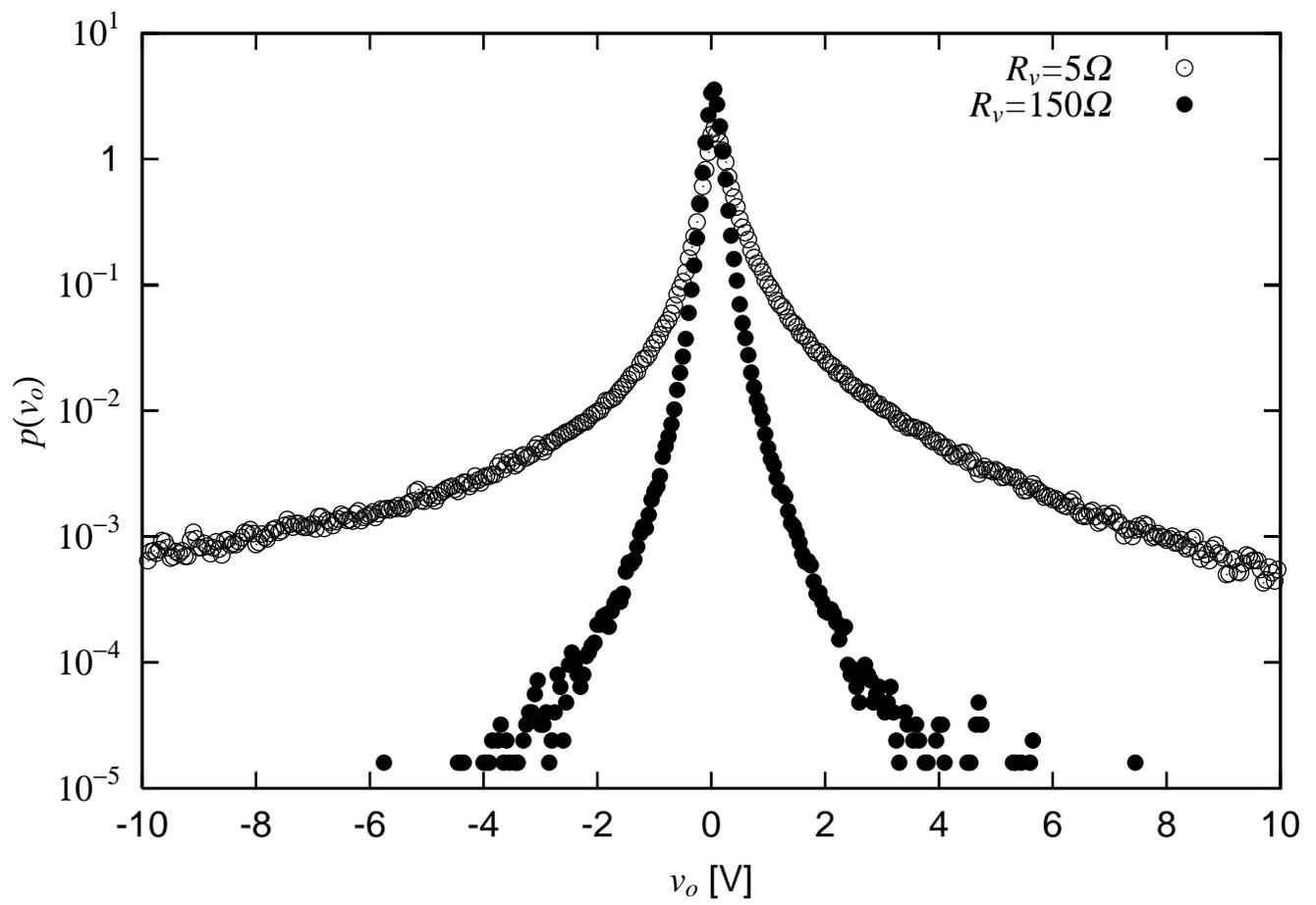
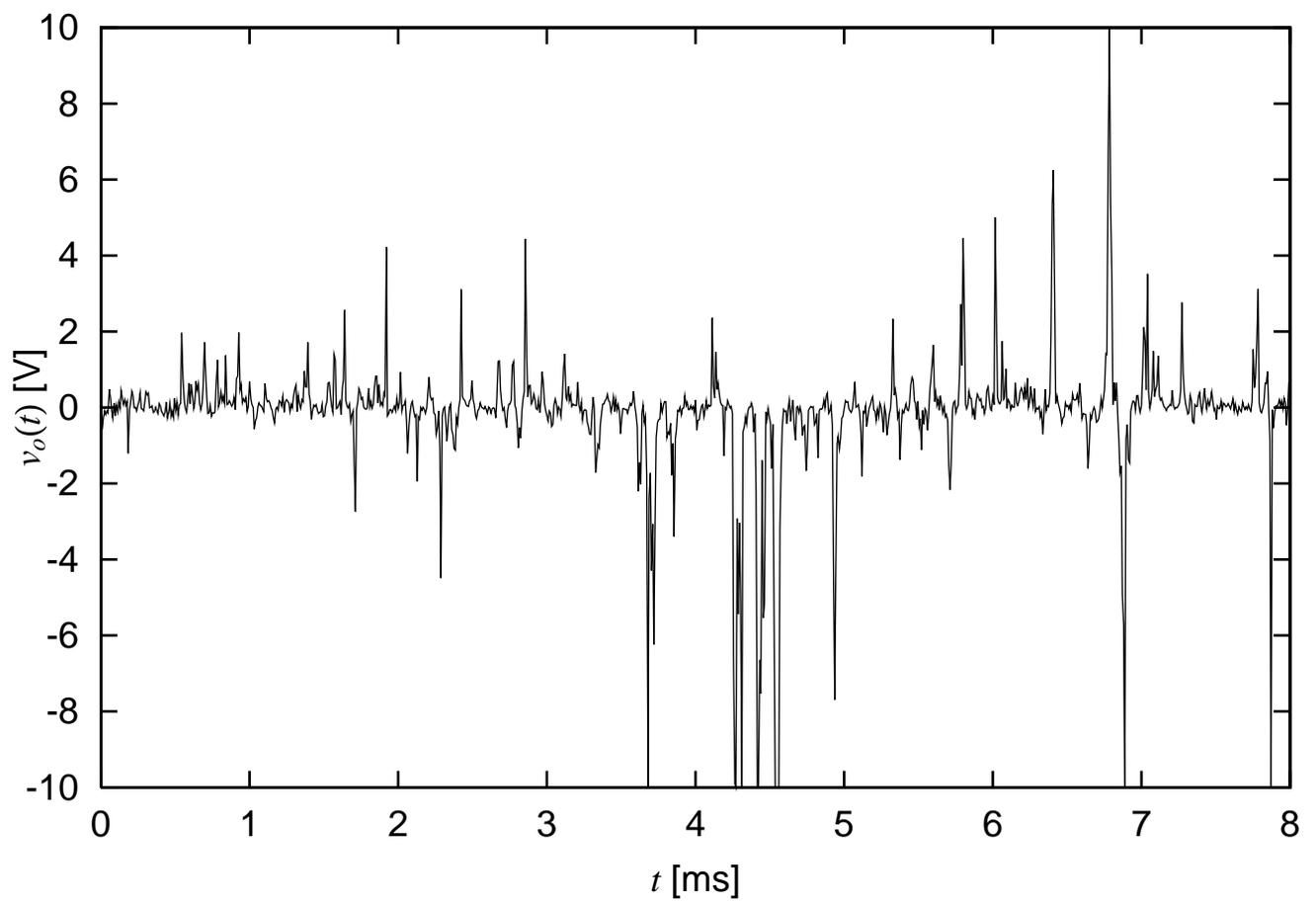
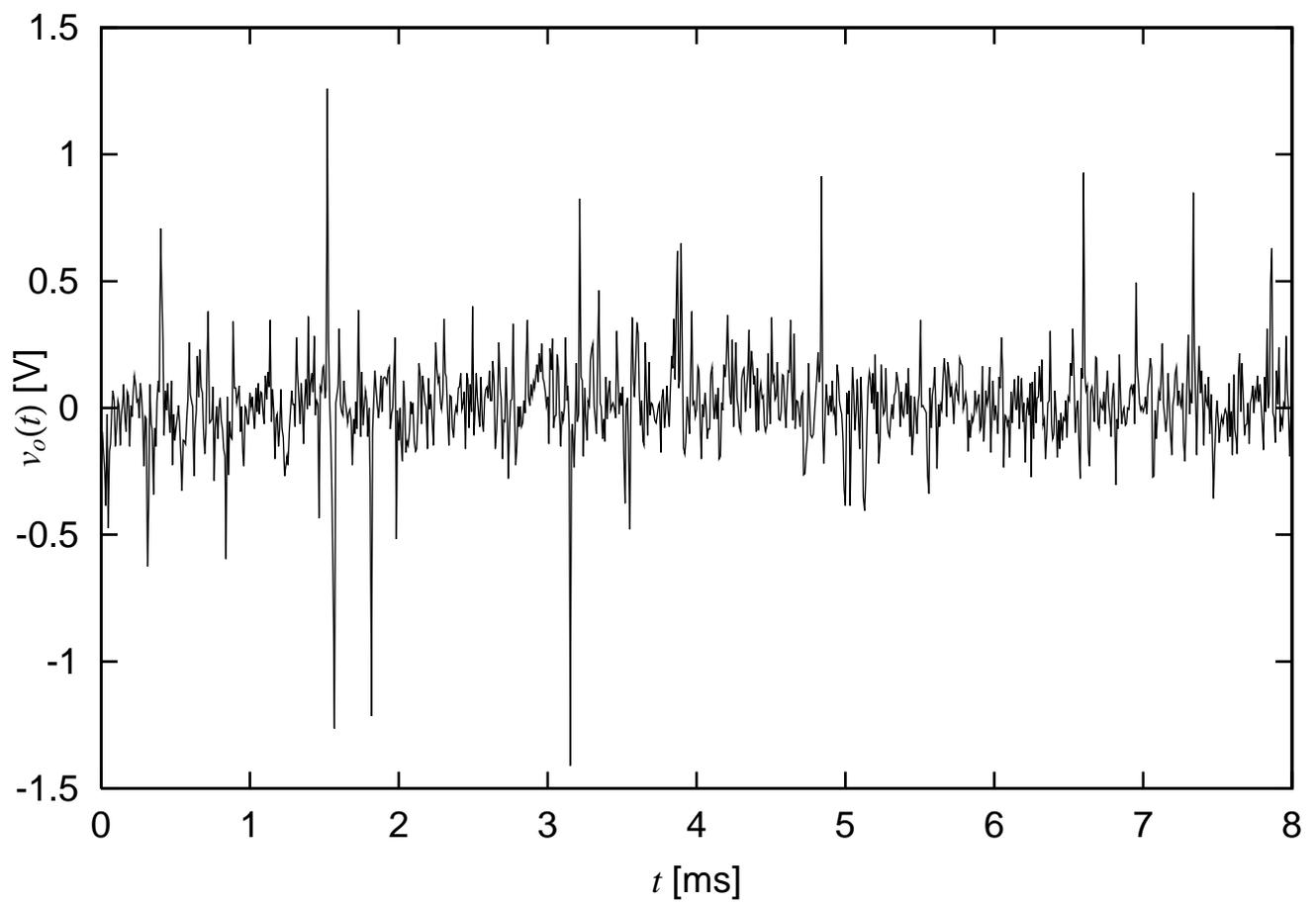
\end{document}